\documentclass[review]{elsarticle}
\usepackage[top=40mm, bottom=35mm, left=25mm, right=30mm]{geometry}
\usepackage{lineno,hyperref}
\usepackage{color}
\usepackage{float}
\usepackage{array}
\usepackage{amsmath}
\usepackage{stmaryrd}
\usepackage{stackrel}
\usepackage{graphicx}
\usepackage{setspace}
\usepackage{esint}
\usepackage{bbold}
\modulolinenumbers[5]

\journal{Journal of \LaTeX\ Templates}

\begin{document}

\begin{frontmatter}

\title{Analytical solutions of the QCD$\otimes$QED DGLAP evolution equations based on the Mellin transform technique in NLO approximation}


\author[mymainaddress]{Marzieh Mottaghizadeh}

\author[mymainaddress]{Fatemeh Taghavi Shahri\corref{mycorrespondingauthor}}
\cortext[mycorrespondingauthor]{Corresponding author}
\ead{taghavishahri@um.ac.ir}

\author[mymainaddress]{Parvin Eslami}

\address[mymainaddress]{Department of Physics, Ferdowsi University of Mashhad,
Mashhad, Iran}

\begin{abstract}

In this paper we present a new and efficient analytical solutions for evolving the QCD$\otimes$QED
DGLAP evolution equations in mellin space and obtain the parton distribution functions
(PDFs) in perturbative QCD including the QED corrections. The validity of our analytical solutions, which have done in the next to leading order QCD and the leading order QED approximations, are checked with the initial parton distributions from newly released CT14QED global analysis code (Phys. Rev.D93,114015(2016)). The evolved parton distribution functions are in good agreement with  CT14QED PDFs set and also with those from APFEL (Computer Physics Communications 185, 1647 (2014)) program. Finally, we derived the impact of the NLO QED corrections to the QCD$\otimes$QED DGLAP evolution equations.

\end{abstract}

\begin{keyword}
\texttt{QCD$\otimes$QED DGLAP evolution equations\sep Mellin moment \sep Parton distribution functions.}

\end{keyword}

\end{frontmatter}

\linenumbers

\section{Introduction }

The new search at the LHC demands knowledge of the photon distribution
function at different values of $x$ and $Q^{2}$. The well-known Dokshitzer-Gribov-Lipatov-Altarelli-Parisi (DGLAP) integrate-differential evolution equations give the parton distribution
functions in the pertobative QCD \cite{Altarelli:1977zs,Dokshitzer:1977sg,Gribov:1972ri,Lipatov:1974qm}.
There exists many published papers attempting to give solutions of DGLAP equations in the realm of QCD, mostly based on the global parameterization of PDF's, but only a few have also included corrections  coming from QED.  The MRST group \cite{Martin:1998sq,Martin:2004dh} being the first known group to that effect.

Recently the NNPDF collaboration \cite{Bertone:2013vaa,Ball:2013hta}
and CT14QED group \cite{Schmidt:2015zda} reported the new
results on this issue. A precise knowledge of the parton distribution functions of the proton
is presented in Refs. \cite{Placakyte:2011az,Abramowicz:1900rp}
in order to make predictions for the Standard Model and beyond the Standard Model processes at hadron colliders. Recently, Florian et al. \cite{deFlorian:2015ujt}
have extended the available knowledge of the Altarelli-Parisi splitting
functions to one order higher in QED. They have provided expressions
for the splitting kernels up to $O(\alpha\alpha_{S})$. Also, Florian et al. \cite{deFlorian:2016gvk} have computed the two-loop QED corrections to the Altarell-Parisi splitting functions by using a deconstructive algorithmic Abelianization of the NLO QCD corrections.

In the CT14QED global parametrization, the photon parton distribution
function is explained by a two-parameter ansatz, coming from radiation
off the valence quarks, and based on the CT14 next to leading order
(NLO) PDFs. The APFEL global analysis code facilitates the determination of parton distribution
functions with electroweak corrections. Its uncertainties extracted
from the DIS and LHC hadronic data using the Monte Carlo approach adopted
by the Neural Network PDF (NNPDF) methodology.

Recently, F. Giuli and XFitter Developer team ~\cite{Giuli:2017oii} presented a determination of the photon PDF from fits to recent ATLAS measurments of high-mass Drell-Yan dilepton production at $\sqrt{s}=8 TeV$.
Also, the photon distribution function of the proton, $ x \gamma(x, Q^2)$, calculated in terms of inclusive lepton-proton deep-inelastic scattering (DIS) structure functions. The photon distribution function resulting from this strategy has been realized in LUXQED PDFs set ~\cite{Manohar:2016nzj}, which is in the next to next leading order (NNlo) QCD and NLO QED approximations.

The DGLAP evolution equations can be solved
numerically by different methods such as the Mellin and Laplace transform technique \cite{Moch:2004pa,Vogt:2004ns,Block:2007pg,Block:2010du,Mottaghizadeh:2016krr,TaghaviShahri:2010be,AtashbarTehrani:2013qea,Zarei:2015jvh,Zarrin:2016kxf}.
The aim of this paper is to investigate the analytical solutions of
the QCD$\otimes$QED DGLAP evolution equations and present an approach
based on the Mellin transform technique. The present theoretical abilities allow
extensive calculations at the higher order corrections of QCD and
QED.

This article is organized as follows: in section 2,
we analytically solved the QCD$\otimes$QED DGLAP evolution equations
in N-moment space. We also investigated the NLO QED corrections to the QCD$\otimes$QED DGLAP evolution equations in Mellin space in section 3. In section 4, we tested the validity  of solutions and compared our theoretical predictions of the parton distribution functions with those from global analysis codes. Finally,
section 5 is devoted to the conclusions.
\section{Evolution Equations}
The singlet parton distribution functions $f_{i}(x,Q^{2})$ obey the
DGLAP evolution equations with QED corrections \cite{Carrazza:2015dea}
in x space, as
\begin{equation}
\frac{\partial}{\partial logQ^{2}}\left(\begin{array}{c}
f_{1}\\
f_{2}\\
f_{3}\\
f_{4}
\end{array}\right)=\\
\left(\begin{array}{cccc}
P_{11} & P_{12} & P_{13} & P_{14}\\
P_{21} & P_{22} & P_{23} & P_{24}\\
P_{31} & P_{32} & P_{33} & P_{34}\\
P_{41} & P_{42} & P_{43} & P_{44}
\end{array}\right)\otimes\left(\begin{array}{c}
f_{1}\\
f_{2}\\
f_{3}\\
f_{4}
\end{array}\right)\label{eq:1}
\end{equation}
and the DGLAP evolution equations with QED corrections for the non-singlet parton distribution
functions are as follows,
\begin{equation}
\frac{\partial{{f}_{i}}}{\partial log{{Q}^{2}}}={{P}_{ii}}\otimes{{f}_{i}}\qquad i=5,\ldots,9\label{eq:2}
\end{equation}
where $P_{ij}$ and $P_{ii}$ are the splitting functions and are given
in Ref.\cite{Mottaghizadeh:2016krr} with details,
and $\otimes$ denotes the convolution integral
\begin{equation}
f\otimes g=\intop_{x}^{1}\frac{dy}{y}f(y)g(\frac{x}{y})\label{eq:3}
\end{equation}
We utilize a PDF basis for the QCD$\otimes$QED
DGLAP evolution equations. This basis define by the following singlet and non-singlet
PDF combinations \cite{Roth:2004ti},
\begin{equation}
q^{SG}:\left(\begin{array}{c}
{{f}_{1}}=\Delta=\\
u+\bar{u}+c+\bar{c}-d-\bar{d}-s-\bar{s}-b-\bar{b}\\
{{f}_{2}}=\Sigma=\\
u+\bar{u}+c+\bar{c}+d+\bar{d}+s+\bar{s}+b+\bar{b}\\
{{f}_{3}}=g\\
{{f}_{4}}=\gamma
\end{array}\right)\label{eq:4}
\end{equation}
\begin{center}
\begin{equation}
q^{NS}:\left(\begin{array}{c}
{{f}_{5}}={{d}_{v}}=d-\bar{d}\\
{{f}_{6}}={{u}_{v}}=u-\bar{u}\\
{{f}_{7}}={{\Delta}_{ds}}=d+\bar{d}-s-\bar{s}\\
{{f}_{8}}={{\Delta}_{uc}}=u+\bar{u}-c-\bar{c}\\
{{f}_{9}}={{\Delta}_{sb}}=s+\bar{s}-b-\bar{b}
\end{array}\right)\label{eq:5}
\end{equation}
\par\end{center}
Here in the next subsections we bring out the solutions of the QCD$\otimes$QED
DGLAP evolution equations with more details. Our solutions is done
in the next to leading order QCD and the leading order QED approximations.

\subsection{The singlet PDFs with LO QED corrections}

In the singlet sector, $f$ in equation ~\eqref{eq:4} is a four component vector. The Mellin transform is defined as,

\begin{equation}
f^{N}=\stackrel[0]{1}{\int dx}x^{N-1}f(x)\label{eq:6}
\end{equation}
Therefore, the evolution equation ~\eqref{eq:1}
factorizes into the form
\begin{equation}
\frac{\partial f^{N}(Q^{2})}{\partial logQ^{2}}=P^{N}(Q^{2})\cdot f^{N}(Q^{2})\label{eq:7}
\end{equation}
\textcolor{black}{where $P^{N}(Q^{2})$ is the $4\times4$ splitting
matrix.}
We can expand this splitting matrix in the next-to-leading order QCD
and the leading order QED approximations as follows,
\begin{equation}
P^{N}(Q^{2})=a_{s}(Q^{2})P_{0}^{(1,0)}(N)+a_{s}^{2}(Q^{2})P^{(2,0)}(N)\\
+a(Q^{2})P_{0}^{(0,1)}(N)\label{eq:8}
\end{equation}
where $a_{s}\equiv\frac{\alpha_{s}}{4\pi}$ and $a\equiv\frac{\alpha}{4\pi}$
are the QCD and QED running couplings, respectively.
We can separated the splitting matrix into two parts of QCD and QED.
Then we have,
\begin{eqnarray*}
P_{QCD}^{N}(Q^{2}) & = & a_{s}(Q^{2})P_{0}^{(1,0)}(N)+a_{s}^{2}(Q^{2})P^{(2,0)}(N) +\mathcal{O}(a_{s}^{3}),\\
P_{QED}^{N}(Q^{2}) & = & a(Q^{2})P_{0}^{(0,1)}(N)+\mathcal{O}(a^{2}).
\end{eqnarray*}
The QED splitting matrix, $P_{0}^{(0,1)}(N)$, and the QCD splitting matrices $P_{0}^{(1,0)}(N)$
and $P^{(2,0)}(N)$ are represented as follows,
\begin{center}
$P_{0}^{(1,0)}(N)=\left(\begin{array}{cccc}
p_{qq}^{(0)} & 0 & \frac{n_{u}-n_{d}}{n_{f}}p_{qg}^{(0)} & 0\\
0 & p_{qq}^{(0)} & p_{qg}^{(0)} & 0\\
0 & p_{gq}^{(0)} & p_{gg}^{(0)} & 0\\
0 & 0 & 0 & 0
\end{array}\right)$,
\par\end{center}

\[
P^{(2,0)}(N)=\left(\begin{array}{cccc}
p_{+}^{(1)} & \frac{n_{u}-n_{d}}{n_{f}}(p_{qq}^{(1)}-p_{+}^{(1)}) & p_{qg}^{(1)} & 0\\
0 & p_{qq}^{(1)} & p_{qg}^{(1)} & 0\\
0 & p_{gq}^{(1)} & p_{gg}^{(1)} & 0\\
0 & 0 & 0 & 0
\end{array}\right),
\]

\begin{equation}
P_{0}^{(0,1)}(N)=\\
\left(\begin{array}{cccc}
\eta^+\tilde{p}_{qq}^{(0)} & \eta^-\tilde{p}_{qq}^{(0)} & 0 & \delta_{e}^{2} p_{q\gamma}^{(0)}\\
\eta^-\tilde{p}_{qq}^{(0)} & \eta^+\tilde{p}_{qq}^{(0)} & 0 & e_{\varSigma}^{2} p_{q\gamma}^{(0)}\\
0 & 0 & 0 & 0\\
\eta^- p_{\gamma q}^{(0)} & \eta^+ p_{\gamma q}^{(0)} & 0 & p_{\gamma\gamma}^{(0)}
\end{array}\right)
\end{equation}
where 
\begin{align*}
e_{\varSigma}^{2} & \equiv N_{c}(\frac{n_{u}e_{u}^{2}+n_{d}e_{d}^{2}}{n_{f}})\\
\delta_{e}^{2} & \equiv N_{c}(\frac{n_{u}e_{u}^{2}-n_{d}e_{d}^{2}}{n_{f}})\\
\eta^{\pm} & \equiv\frac{1}{2}(e_u^2\pm e_d^2)
\end{align*}
The $n_{u}$ and $n_{d}$ parameters are the number of up and down-type active
quark flavors, respectively, and $n_{f}=n_{u}+n_{d}$, where, $n_{f}$ is the number of active flavors. The moments of Altraelli-Parisi function may be found in Refs. \cite{Roth:2004ti,Floratos:1978ny,Floratos:1977au,GonzalezArroyo:1979df,GonzalezArroyo:1979he,Floratos:1981hs,Furmanski:1981cw,Curci:1980uw,Hamberg:1991qt,Moch:1999eb}, noting that $P_{i}^{z}=+\gamma^{(i)z}/2$.
The running couplings are given by
\begin{eqnarray}
a_{s}(Q^{2}) & =\frac{1}{\beta_{0}^{(\alpha_s)}log(\frac{Q^{2}}{\Lambda_{QCD}^{2}})}(1-\frac{\beta_{1}^{(\alpha_s^2)}}{\beta_{0}^{(\alpha_s)^{2}}}\frac{log(log(\frac{Q^{2}}{\Lambda_{QCD}^{2}}))}{log(\frac{Q^{2}}{\Lambda_{QCD}^{2}})})\label{eq:9}
\end{eqnarray}
\begin{eqnarray}
a(Q^{2}) & =\frac{a(\mu^{2})}{1+\beta_{0}^{(\alpha)}a(\mu^{2})log(\frac{Q^{2}}{\mu^{2}})}\label{eq:10}
\end{eqnarray}
where, $\beta_{0}^{(\alpha_s)}=\frac{1}{3}(33-2n_{f})$ and $\beta_{1}=102-\frac{38}{3}n_{f}$.
\\The lowest-order QED beta function is
\begin{equation}
\beta_{0}^{(\alpha)}=-\frac{4}{3}(n_{l}+N_{c}\stackrel[i=1]{n_{f}}{\sum}e_{i}^{2})
\end{equation}
where, $e_{i}$ is the relative electric charge of the quarks and
$N_{c}$ denotes the multiplicity due to color degrees of freedom,
i.e.,$N_{c}=3$ for quarks. The $n_{l}$ parameter is the number of active lepton flavour.

The evolution equations of  these couplings are given by the QCD
and QED $\beta$-functions as
\[
Q^{2}\frac{\partial a_{s}}{\partial Q^{2}}=\beta_{QCD}(a_{s}),\qquad Q^{2}\frac{\partial a}{\partial Q^{2}}=\beta_{QED}(a),
\]
where, the beta functions for the strong and electromagnetic couplings
are expanded to the appropriate order, as
\begin{equation}
\beta_{QCD}(a_{s})=-\beta_{0}^{(\alpha_s)}{(\alpha_s)}a_{s}^{2}-\beta_{1}^{(\alpha_s^2)}a_{s}^{3}+\mathcal{O}(a_{s}^{4}),\label{eq:11}
\end{equation}
\begin{equation}
\beta_{QED}(a)=-\beta_{0}^{(\alpha)}a^{2}+\mathcal{O}(a^{3}),\label{eq:12}
\end{equation}
We obtain the solution of the equation ~\eqref{eq:7}
in a compact form,
\begin{equation}
f(N,Q^{2})=E^{N}(Q^{2},Q_{0}^{2})f(N,Q_{0}^{2})\label{eq:13}
\end{equation}
where $E^{N}(Q^{2},Q_{0}^{2})$ is the evolution matrix, as
\begin{equation}
E^{N}(Q^{2},Q_{0}^{2})=E_{QCD}^{N}(Q^{2},Q_{0}^{2})\cdot E_{QED}^{N}(Q^{2},Q_{0}^{2})\label{eq:14}
\end{equation}
where $E_{QCD}^{N}(Q^{2},Q_{0}^{2})$ and $E_{QED}^{N}(Q^{2},Q_{0}^{2})$,
are the QCD and QED evolution matrices, respectively. 
The QCD evolution matrix, $E_{QCD}^{N}(Q^{2},Q_{0}^{2})$, is the
solution of following equation
\begin{equation}
\frac{\partial E_{QCD}^{N}}{\partial logQ^{2}}=a_{s}(Q^{2})P_{0}^{(1,0)}(N)\cdot E_{QCD}^{N}\label{eq:15}
\end{equation}
Defining t as,
\begin{equation}
t=\frac{1}{\beta_{0}^{(\alpha_s)}} log\frac{a_{s}(Q_{0}^{2})}{a_{s}(Q^{2})}\label{eq:16}
\end{equation}
The equation. ~\eqref{eq:15} can be write
as follows,
\begin{equation}
\frac{d}{dt}E_{QCD}(t)=(P_{0}^{(1,0)}(N)+\frac{\alpha_{s}}{4\pi}R+\mathcal{O}(\alpha_{s}^{2}))E_{QCD}(t)\label{eq:17}
\end{equation}
where, $R=P^{(2,0)}(N)-\frac{\beta_{1}^{(\alpha_s^2)}}{\beta_{0}^{(\alpha_s)}}P_{0}^{(1,0)}(N)$.
Let us denote the leading log solution of equation ~\eqref{eq:17}
by $E_{QCD}^{(0)}(t)$,i.e.,
\begin{equation}
\frac{d}{dt}E_{QCD}^{(0)}(t)=P_{0}^{(1,0)}(N)E_{QCD}^{(0)}(t)\label{eq:18}
\end{equation}
Then the full solution of equation ~\eqref{eq:17}
can be write as a power series of $\alpha_{s}$ as,
\begin{equation}
E_{QCD}(t)=(1+\frac{\alpha_{s}}{4\pi}U+\mathcal{O}(\alpha_{s}^{2}))E_{QCD}^{(0)}(t)\label{eq:19}
\end{equation}
Substituting the equation ~\eqref{eq:19} into equation
~\eqref{eq:17}, we obtain the following equation
for the new evolution matrix $U$:
\begin{equation}
\left[U,P_{0}^{(1,0)}(N)\right]=\beta_{0}^{(\alpha_s)}U+R\label{eq:20}
\end{equation}
where, $\left[U,P_{0}^{(1,0)}(N)\right]=UP_{0}^{(1,0)}(N)-P_{0}^{(1,0)}(N)U$.
In the QCD singlet case, the equation ~\eqref{eq:20}
is less trivial, since $\left[U,P_{0}^{(1,0)}(N)\right]$ does not vanish
in general. Usually, one solves equation ~\eqref{eq:20}
by going to a frame where the matrix $P_{0}^{(1,0)}(N)$ is diagonal.
It appears that it is then rather difficult to write the final solution
in a compact matrix form. Here we present a rather simple matrix solution
of equation ~\eqref{eq:17} in the singlet case.
It should be noted that all of the calculations are done in Mellin space.
The eigenvalues of the $4\times4$ matrix of $P_{0}^{(1,0)}(N)$, are
given by

\[
\lambda_{1,2}=\frac{1}{2}[p_{qq}^{(0)}+p_{gg}^{(0)}\pm\sqrt{(p_{qq}^{(0)}-p_{gg}^{(0)})^{2}+4p_{qg}^{(0)}p_{gq}^{(0)}}],
\]
\[
\lambda_{3}=p_{qq}^{(0)}\qquad and\qquad\lambda_{4}=0.
\]
Using the projection operators, we can write-down $P_{0}^{(1,0)}(N)$
as
\begin{equation}
P_{0}^{(1,0)}(N)=\lambda_{1}e_{1}+\lambda_{2}e_{2}+\lambda_{3}e_{3}+\lambda_{4}e_{4}\label{eq:21}
\end{equation}
where, the projection operators are given by
\begin{center}
$e1\equiv\left(\begin{array}{cccc}
0 & -\frac{\frac{n_{u}-n_{d}}{n_{f}}p_{gq}^{(0)}p_{qg}^{(0)}}{(\lambda_{2}-\lambda_{1})(\lambda_{1}-p_{qq}^{(0)})} & -\frac{\frac{n_{u}-n_{d}}{n_{f}}p_{qg}^{(0)}}{(\lambda_{2}-\lambda_{1})} & 0\\
0 & -\frac{p_{qg}^{(0)}p_{gq}^{(0)}}{(\lambda_{2}-\lambda_{1})(\lambda_{1}-p_{qq}^{(0)})} & -\frac{p_{qg}^{(0)}}{(\lambda_{2}-\lambda_{1})} & 0\\
0 & -\frac{p_{gq}^{(0)}}{(\lambda_{2}-\lambda_{1})} & -\frac{(\lambda_{1}-p_{qq}^{(0)})}{(\lambda_{2}-\lambda_{1})} & 0\\
0 & 0 & 0 & 0
\end{array}\right)$,
\par\end{center}

\begin{center}
$e2\equiv\left(\begin{array}{cccc}
0 & \frac{\frac{n_{u}-n_{d}}{n_{f}}p_{gq}^{(0)}p_{qg}^{(0)}}{(\lambda_{2}-\lambda_{1})(\lambda_{2}-p_{qq}^{(0)})} & \frac{\frac{n_{u}-n_{d}}{n_{f}}p_{qg}^{(0)}}{(\lambda_{2}-\lambda_{1})} & 0\\
0 & \frac{p_{qg}^{(0)}p_{gq}^{(0)}}{(\lambda_{2}-\lambda_{1})(\lambda_{2}-p_{qq}^{(0)})} & \frac{p_{qg}^{(0)}}{(\lambda_{2}-\lambda_{1})} & 0\\
0 & \frac{p_{gq}^{(0)}}{(\lambda_{2}-\lambda_{1})} & \frac{(\lambda_{2}-p_{qq}^{(0)})}{(\lambda_{2}-\lambda_{1})} & 0\\
0 & 0 & 0 & 0
\end{array}\right)$,
\par\end{center}

\begin{center}
$e_{3}\equiv\left(\begin{array}{cccc}
1 & -\frac{n_{u}-n_{d}}{n_{f}} & 0 & 0\\
0 & 0 & 0 & 0\\
0 & 0 & 0 & 0\\
0 & 0 & 0 & 0
\end{array}\right)$,
\par\end{center}

\begin{center}
$e4\equiv\left(\begin{array}{cccc}
0 & 0 & 0 & 0\\
0 & 0 & 0 & 0\\
0 & 0 & 0 & 0\\
0 & 0 & 0 & 1
\end{array}\right)$.
\par\end{center}
We have
\begin{eqnarray}
e_{i}^{2}=e_{i},\quad i=1,\ldots,4 \nonumber\\
e_{i}e_{j}=0,\quad i=1,\ldots,4,j=1,\ldots,4,i\neq j \nonumber\\
e_{1}+e_{2}+e_{3}+e_{4}=\mathbb{1}
\end{eqnarray}
where, $\mathbb{1}$ is the unit $4\times4$ matrix.
The solution of the equation ~\eqref{eq:18} can
now be written as
\begin{equation}
E^{(0)}(t)=e^{P_{0}^{(1,0)}(N)t}=\\
e_{1}e^{\lambda_{1}t}+e_{2}e^{\lambda_{2}t}+e_{3}e^{\lambda_{3}t}+e_{4}e^{\lambda_{4}t}\label{eq:22}
\end{equation}
Since $e_{1}+e_{2}+e_{3}+e_{4}=\mathbb{1}$, we have an obvious identity matrix
as,
\begin{eqnarray}
U=e_{1}Ue_{1}+e_{1}Ue_{2}+e_{1}Ue_{3}+e_{1}Ue_{4}\cr
+e_{2}Ue_{1}+e_{2}Ue_{2}+e_{2}Ue_{3}+e_{2}Ue_{4}\cr
+e_{3}Ue_{1}+e_{3}Ue_{2}+e_{3}Ue_{3}+e_{3}Ue_{4}\cr
+ e_{4}Ue_{1}+e_{4}Ue_{2}+e_{4}Ue_{3}+e_{4}Ue_{4}\label{eq:23}
\end{eqnarray}
Inserting the Eq. ~\eqref{eq:21} into Eq. ~\eqref{eq:20}, yields
\begin{eqnarray}
U=-\frac{1}{\beta_{0}^{(\alpha_s)}}(e_{1}Re_{1}+e_{2}Re_{2}+e_{3}Re_{3}+e_{4}Re_{4})\cr
+\frac{e_{1}Re_{2}}{\lambda_{2}-\lambda_{1}-\beta_{0}^{(\alpha_s)}}+\frac{e_{1}Re_{3}}{\lambda_{1}-\lambda_{2}-\beta_{0}^{(\alpha_s)}}+\frac{e_{1}Re_{4}}{\lambda_{4}-\lambda_{1}-\beta_{0}^{(\alpha_s)}}\cr
+\frac{e_{2}Re_{1}}{\lambda_{1}-\lambda_{2}-\beta_{0}^{(\alpha_s)}}+\frac{e_{2}Re_{3}}{\lambda_{3}-\lambda_{2}-\beta_{0}^{(\alpha_s)}}+\frac{e_{2}Re_{4}}{\lambda_{4}-\lambda_{2}-\beta_{0}^{(\alpha_s)}}\cr
+\frac{e_{3}Re_{1}}{\lambda_{1}-\lambda_{3}-\beta_{0}^{(\alpha_s)}}+\frac{e_{3}Re_{2}}{\lambda_{2}-\lambda_{3}-\beta_{0}^{(\alpha_s)}}+\frac{e_{3}Re_{4}}{\lambda_{4}-\lambda_{3}-\beta_{0}^{(\alpha_s)}}\cr
+\frac{e_{4}Re_{1}}{\lambda_{1}-\lambda_{4}-\beta_{0}^{(\alpha_s)}}+\frac{e_{4}Re_{2}}{\lambda_{2}-\lambda_{4}-\beta_{0}^{(\alpha_s)}}+\frac{e_{4}Re_{3}}{\lambda_{3}-\lambda_{4}-\beta_{0}^{(\alpha_s)}}\label{eq:24}
\end{eqnarray}
Therefore, The final solution for the QCD evolution matrix, $E_{QCD}^{N}(t)$,
in the next to leading order approximation can be obtained from the equation
~\eqref{eq:19} as
\begin{eqnarray}
E_{QCD}^{N}(a_{s},a_{s0})=\{(\frac{a_{s}}{a_{s0}})^{\frac{\lambda_{1}}{\text{\ensuremath{\beta_{0}^{(\alpha_s)}}}}}(e_{1}-\frac{(\text{\ensuremath{a_{s0}-a_{s})}}}{\text{\ensuremath{\beta_{0}^{(\alpha_s)}}}}e_{1}Re_{1}-\frac{1}{\beta_{0}^{(\alpha_s)}+\lambda_{2}-\lambda_{1}}(a_{s}(\frac{a_{s}}{a_{s0}})^{(\lambda_{2}-\lambda_{1})/\beta_{0}^{(\alpha_s)}}-a_{s0})e_{1}Re_{2}\cr
-\frac{1}{\beta_{0}^{(\alpha_s)}+\lambda_{3}-\lambda_{1}}(a_{s}(\frac{a_{s}}{a_{s0}})^{(\lambda_{3}-\lambda_{1})/\beta_{0}^{(\alpha_s)}}-a_{s0})e_{1}Re_{3}-\frac{1}{\beta_{0}^{(\alpha_s)}+\lambda_{4}-\lambda_{1}}(a_{s}(\frac{a_{s}}{a_{s0}})^{(\lambda_{4}-\lambda_{1})/\beta_{0}^{(\alpha_s)}}-a_{s0})e_{1}Re_{4})\cr
+(\frac{a_{s}}{a_{s0}})^{\frac{\lambda_{2}}{\text{\ensuremath{\beta_{0}^{(\alpha_s)}}}}}(e_{2}-\frac{(\text{\ensuremath{a_{s0}-a_{s})}}}{\text{\ensuremath{\beta_{0}^{(\alpha_s)}}}}e_{2}Re_{2}-\frac{1}{\beta_{0}^{(\alpha_s)}+\lambda_{1}-\lambda_{2}}(a_{s}(\frac{a_{s}}{a_{s0}})^{(\lambda_{1}-\lambda_{2})/\beta_{0}^{(\alpha_s)}}-a_{s0})e_{2}Re_{1}\cr
-\frac{1}{\beta_{0}^{(\alpha_s)}+\lambda_{3}-\lambda_{2}}(a_{s}(\frac{a_{s}}{a_{s0}})^{(\lambda_{3}-\lambda_{2})/\beta_{0}^{(\alpha_s)}}-a_{s0})e_{2}Re_{3}-\frac{1}{\beta_{0}^{(\alpha_s)}+\lambda_{4}-\lambda_{2}}(a_{s}(\frac{a_{s}}{a_{s0}})^{(\lambda_{4}-\lambda_{2})/\beta_{0}^{(\alpha_s)}}-a_{s0})e_{2}Re_{4})\cr
+(\frac{a_{s}}{a_{s0}})^{\frac{\lambda_{3}}{\text{\ensuremath{\beta_{0}^{(\alpha_s)}}}}}(e_{3}-\frac{(\text{\ensuremath{a_{s0}-a_{s})}}}{\text{\ensuremath{\beta_{0}^{(\alpha_s)}}}}e_{3}Re_{3}-\frac{1}{\beta_{0}^{(\alpha_s)}+\lambda_{1}-\lambda_{3}}(a_{s}(\frac{a_{s}}{a_{s0}})^{(\lambda_{1}-\lambda_{3})/\beta_{0}^{(\alpha_s)}}-a_{s0})e_{3}Re_{1}\cr
-\frac{1}{\beta_{0}^{(\alpha_s)}+\lambda_{2}-\lambda_{3}}(a_{s}(\frac{a_{s}}{a_{s0}})^{(\lambda_{2}-\lambda_{3})/\beta_{0}^{(\alpha_s)}}-a_{s0})e_{3}Re_{2}-\frac{1}{\beta_{0}^{(\alpha_s)}+\lambda_{4}-\lambda_{3}}(a_{s}(\frac{a_{s}}{a_{s0}})^{(\lambda_{4}-\lambda_{3})/\beta_{0}^{(\alpha_s)}}-a_{s0})e_{3}Re_{4})\cr
+(\frac{a_{s}}{a_{s0}})^{\frac{\lambda_{4}}{\text{\ensuremath{\beta_{0}^{(\alpha_s)}}}}}(e_{4}-\frac{(\text{\ensuremath{a_{s0}-a_{s})}}}{\text{\ensuremath{\beta_{0}^{(\alpha_s)}}}}e_{4}Re_{4}-\frac{1}{\beta_{0}^{(\alpha_s)}+\lambda_{1}-\lambda_{4}}(a_{s}(\frac{a_{s}}{a_{s0}})^{(\lambda_{1}-\lambda_{4})/\beta_{0}^{(\alpha_s)}}-a_{s0})e_{4}Re_{1}\cr
-\frac{1}{\beta_{0}^{(\alpha_s)}+\lambda_{2}-\lambda_{4}}(a_{s}(\frac{a_{s}}{a_{s0}})^{(\lambda_{2}-\lambda_{4})/\beta_{0}^{(\alpha_s)}}-a_{s0})e_{4}Re_{2}-\frac{1}{\beta_{0}^{(\alpha_s)}+\lambda_{3}-\lambda_{4}}(a_{s}(\frac{a_{s}}{a_{s0}})^{(\lambda_{3}-\lambda_{4})/\beta_{0}^{(\alpha_s)}}-a_{0s})e_{4}Re_{3})\}\label{eq:25}
\end{eqnarray}
where, $a_{s0}$ is $a_{s}$ at a initial scale of $Q_{0}^{2}$.
\\
Now, the leading order QED evolution matrix, $E_{QED}^{N}(a,a_{0})$ in terms of $a$ and $a_{0}$ at the scales of $Q^{2}$ and $Q_{0}^{2}$, respectively, can be represented as

\begin{equation}
E_{QED}^{N}(a,a_{0})=\tilde{e}_{1}(\frac{a}{a_{0}})^{\frac{\tilde{\lambda}_{1}}{\text{\ensuremath{\beta_{0}^{(\alpha)}}}}}+\tilde{e}_{2}(\frac{a}{a_{0}})^{\frac{\tilde{\lambda}_{2}}{\text{\ensuremath{\beta_{0}^{(\alpha)}}}}}+\tilde{e}_{3}(\frac{a}{a_{0}})^{\frac{\tilde{\lambda}_{3}}{\beta_{0}^{(\alpha)}}}+\tilde{e}_{4}(\frac{a}{a_{0}})^{\frac{\tilde{\lambda}_{4}}{\text{\ensuremath{\beta_{0}^{(\alpha)}}}}}\label{eq:26}
\end{equation}
The matrices $\tilde{e}_{1}$, $\tilde{e}_{2}$, $\tilde{e}_{3}$
and $\tilde{e}_{4}$ denote the corresponding projectors,
\begin{center}
$\tilde{e}_{1}=-\frac{-(P_{0}^{(0,1)}(N))^{2}+P_{0}^{(0,1)}(N)(\tilde{\lambda}_{2}+\tilde{\lambda}_{3})-\tilde{\lambda}_{2}\tilde{\lambda}_{3}\mathbb{1}+\tilde{e}_{4}\tilde{\lambda}_{2}\tilde{\lambda}_{3}}{(\tilde{\lambda}_{1}-\tilde{\lambda}_{2})(\tilde{\lambda}_{1}-\tilde{\lambda}_{3})}$
\par\end{center}

\begin{center}
$\tilde{e}_{2}=-\frac{-(P_{0}^{(0,1)}(N))^{2}+P_{0}^{(0,1)}(N)(\tilde{\lambda}_{1}+\tilde{\lambda}_{3})-\tilde{\lambda}_{1}\tilde{\lambda}_{3}\mathbb{1}+\tilde{e}_{4}\tilde{\lambda}_{1}\tilde{\lambda}_{3}}{(-\tilde{\lambda}_{1}+\tilde{\lambda}_{2})(\tilde{\lambda}_{2}-\tilde{\lambda}_{3})}$
\par\end{center}

\begin{center}
$\tilde{e}_{3}=-\frac{(P_{0}^{(0,1)}(N))^{2}-P_{0}^{(0,1)}(N)(\tilde{\lambda}_{1}+\tilde{\lambda}_{2})+\tilde{\lambda}_{1}\tilde{\lambda}_{2}\mathbb{1}-\tilde{e}_{4}\tilde{\lambda}_{1}\tilde{\lambda}_{2}}{(\tilde{\lambda}_{2}-\tilde{\lambda}_{3})(\tilde{\lambda}_{3}-\tilde{\lambda}_{1})}$
\par\end{center}

\begin{center}
$\tilde{e}_{4}\equiv\left(\begin{array}{cccc}
0 & 0 & 0 & 0\\
0 & 0 & 0 & 0\\
0 & 0 & 1 & 0\\
0 & 0 & 0 & 0
\end{array}\right)$.
\par\end{center}
where, $\tilde{\lambda}_{1}$, $\tilde{\lambda}_{2}$, $\tilde{\lambda}_{3}$
and $\tilde{\lambda}_{4}$ are the eigenvalues of QED matrix,
$P_{0}^{(0,1)}(N)$.
\\With the well-known inverse Mellin transform \cite{Graudenz:1995sk}
the parton distribution functions can be derived in x space,
\begin{equation}
f^{k}(x,Q^{2})=\frac{1}{\pi}\int_{0}^{\infty}d\omega Im[e{}^{i\phi}x^{-c-\omega e^{i\phi}}\\
M_{k}(N=c+\omega e^{i\phi},Q^{2})]\label{eq:27}
\end{equation}
where the contour of the integration lies on the right of all singularities
of $M_{k}(N=c+\omega e^{i\phi},Q^{2})$ in the complex N-plane.

\subsection{The non-singlet PDFs with LO QED corrections}

For the non-singlet distribution functions, all of the parameters in
Eq. ~\eqref{eq:2} are scalar. Since the Mellin
transformation turns convolutions into products, the evolution equation
for QCD part, in the next to leading order approximation, becomes
\begin{equation}
\frac{\partial f^{N}(Q^{2})}{\partial logQ^{2}}=(a_{s}(Q^{2})P_{0}^{(1,0)}(N)+a_{s}^{2}(Q^{2})P_{ns}^{(2,0)}(N))f^{N}(Q^{2})\label{eq:28}
\end{equation}
With a change of variable $a_{s}$ instead of $logQ^{2}$ in equation
~\eqref{eq:28}, we have
\begin{equation}
\frac{\partial f^{N}}{\partial a_{s}}=-\frac{P_{0}^{(1,0)}(N)+a_{s}P_{ns}^{(2,0)}(N)}{a_{s}[\beta_{0}^{(\alpha_s)}+\beta_{1}^{(\alpha_s^2)}a_{s}]}f^{N}\label{eq:29}
\end{equation}
Now we expand the right-hand side of equation (\ref{eq:29}) to get
\begin{equation}
\frac{\partial f^{N}}{\partial a_{s}}=-\frac{1}{\beta_{0}^{(\alpha_s)}a_{s}}(P_{0}^{(1,0)}(N)+a_{s}(P_{ns}^{(2,0)}(N)-\frac{\beta_{1}^{(\alpha_s^2)}}{\beta_{0}^{(\alpha_s)}}P_{0}^{(1,0)}(N))f^{N})\label{eq:30}
\end{equation}
Eq. ~\eqref{eq:29} and Eq. ~\eqref{eq:30}
are formally solved by introducing an evolution operator, $E_{QCD}^{N}(Q^{2},Q_{0}^{2})$.
In the non-singlet case this evolution operator is just a scalar function
of N. This evolution operator satisfies the equation ~\eqref{eq:30},
then we have
\begin{equation}
\frac{\partial E_{QCD}^{N}}{\partial a_{s}}=-\frac{1}{\beta_{0}^{(\alpha_s)}a_{s}}(P_{0}^{(1,0)}(N)+a_{s}(P_{ns}^{(2,0)}(N)-\frac{\beta_{1}^{(\alpha_s^2)}}{\beta_{0}^{(\alpha_s)}}P_{0}^{(1,0)}(N)))E_{QCD}^{N}\label{eq:31}
\end{equation}
Solution of this evolution equation is straightforward in the non-singlet
case; expanding beyond the leading terms, we obtain \cite{Blumlein:1996gv,Gluck:1989ze,Gross:1973id,Politzer:1973fx,Jones:1974pg}
\begin{equation}
E_{QCD}^{N}(Q^{2},Q_{0}^{2})=(1+\frac{a_{s}-a_{s0}}{\beta_{0}^{(\alpha_s)}}\\
(P_{ns}^{(2,0)}(N)-\frac{\beta_{1}^{(\alpha_s^2)}}{\beta_{0}^{(\alpha_s)}}P_{0}^{(1,0)}(N)))(\frac{a_{s}}{a_{s0}})^{\frac{P_{0}^{(1,0)}(N)}{\beta_{0}^{(\alpha_s)}}}\label{eq:32}
\end{equation}
With the same procedure we have the QED evolution operator as,
\begin{equation}
E_{QED}^{N}(Q^{2},Q_{0}^{2})=(\frac{a}{a_{0}})^{\frac{P_{0}^{(0,1)}(N)}{\beta_{0}^{(\alpha)}}}\label{eq:33}
\end{equation}
Finally, we simply have the non-singlet PDFs including QED corrections
as
\begin{equation}
f^{N}(Q^{2})=E^{N}(Q^{2},Q_{0}^{2})f^{N}(Q_{0}^{2})\label{eq:34}
\end{equation}
where
\begin{equation}
E^{N}(Q^{2},Q_{0}^{2})=E_{QCD}^{N}(Q^{2},Q_{0}^{2})E_{QED}^{N}(Q^{2},Q_{0}^{2})\label{eq:35}
\end{equation}
The equation ~\eqref{eq:35} has a boundary condition,
$E^{N}(Q_{0}^{2},Q_{0}^{2})=1$, that renders it an entirely perturbation
object. Now by using this method we obtain operator $E^{N}(Q^{2},Q_{0}^{2})$
for all of the distribution functions in equation ~\eqref{eq:2}.
Our results for these functions presented in Table. ~\eqref{table1}.
\begin{table*}
\centering{}\caption{The evolution operators, $E^{N}(Q^{2},Q_{0}^{2})$, for the non-singlet
distribution functions}
\begin{tabular}{c>{\centering}m{11cm}}
\hline
\textbf{$\mathbf{\mathit{f}_{i}}$} & $E^{N}(Q^{2},Q_{0}^{2})$ \tabularnewline
\hline
$f_{5}$ & \begin{singlespace}
\centering{}$(1+\frac{a_{s}-a_{s0}}{\beta_{0}^{(\alpha_s)}}(p_{qq}^{-(2,0)}(N)-\frac{\beta_{1}^{(\alpha_s^2)}}{\beta_{0}^{(\alpha_s)}}p^{(0)}_{qq}(N)))(\frac{a_{s}}{a_{s0}})^{\frac{p^{(0)}_{qq}(N)}{\beta_{0}^{(\alpha_s)}}}(\frac{a}{a_{0}})^{\frac{e_{d}^{2}\tilde{p}_{qq}^{(0)}(N)}{\beta_{0}^{(\alpha)}}}$
\end{singlespace}
\tabularnewline
\hline
$f_{6}$ & \begin{singlespace}
\centering{}$(1+\frac{a_{s}-a_{s0}}{\beta_{0}^{(\alpha_s)}}(p_{qq}^{-(2,0)}(N)-\frac{\beta_{1}^{(\alpha_s^2)}}{\beta_{0}^{(\alpha_s)}}p^{(0)}_{qq}(N)))(\frac{a_{s}}{a_{s0}})^{\frac{p^{(0)}_{qq}(N)}{\beta_{0}^{(\alpha_s)}}}(\frac{a}{a_{0}})^{\frac{e_{u}^{2}\tilde{p}_{qq}^{(0)}(N)}{\beta_{0}^{(\alpha)}}}$
\end{singlespace}
\tabularnewline
\hline
$f_{7}$ & \begin{singlespace}
\centering{}$(1+\frac{a_{s}-a_{s0}}{\beta_{0}^{(\alpha_s)}}(p_{qq}^{+(2,0)}(N)-\frac{\beta_{1}^{(\alpha_s^2)}}{\beta_{0}^{(\alpha_s)}}p^{(0)}_{qq}(N)))(\frac{a_{s}}{a_{s0}})^{\frac{p^{(0)}_{qq}(N)}{\beta_{0}^{(\alpha_s)}}}(\frac{a}{a_{0}})^{\frac{e_{d}^{2}\tilde{p}_{qq}^{(0)}(N)}{\beta_{0}^{(\alpha)}}}$
\end{singlespace}
\tabularnewline
\hline
$f_{8}$ & \begin{singlespace}
\centering{}$(1+\frac{a_{s}-a_{s0}}{\beta_{0}^{(\alpha_s)}}(p_{qq}^{+(2,0)}(N)-\frac{\beta_{1}^{(\alpha_s^2)}}{\beta_{0}^{(\alpha_s)}}p^{(0)}_{qq}(N)))(\frac{a_{s}}{a_{s0}})^{\frac{p^{(0)}_{qq}(N)}{\beta_{0}^{(\alpha_s)}}}(\frac{a}{a_{0}})^{\frac{e_{u}^{2}\tilde{p}_{qq}^{(0)}(N)}{\beta_{0}^{(\alpha)}}}$
\end{singlespace}
\tabularnewline
\hline
$f_{9}$ & \begin{singlespace}
\centering{}$(1+\frac{a_{s}-a_{s0}}{\beta_{0}^{(\alpha_s)}}(p_{qq}^{+(2,0)}(N)-\frac{\beta_{1}^{(\alpha_s^2)}}{\beta_{0}^{(\alpha_s)}}p^{(0)}_{qq}(N)))(\frac{a_{s}}{a_{s0}})^{\frac{p^{(0)}_{qq}(N)}{\beta_{0}^{(\alpha_s)}}}(\frac{a}{a_{0}})^{\frac{e_{d}^{2}\tilde{p}_{qq}^{(0)}(N)}{\beta_{0}^{(\alpha)}}}$
\end{singlespace}
\tabularnewline
\hline
\end{tabular}\label{table1}
\end{table*}

\section{The impact of NLO QED corrections}

In this section, the discussion starts by generalizing the present analytical method for the singlet and non-singlet evolution equations with NLO QED corrections. 

\subsection{The singlet PDFs with NLO QED corrections}

Here, we study the impact of the NLO QED corrections to the singlet DGLAP evolution equations. These equations with NLO QED and QCD corrections include mixed terms of $\alpha_s$ and $\alpha$. Therefore, the QCD and QED $\beta$-function are proportional to $\alpha_s$ and $\alpha$. The evolution equations of couplings $a_s=\frac{\alpha_s}{4 \pi}$ and $a=\frac{\alpha}{4 \pi}$ are given by
\[
Q^{2}\frac{\partial a_{s}}{\partial Q^{2}}=\beta_{QCD}(a_{s},a),\qquad Q^{2}\frac{\partial a}{\partial Q^{2}}=\beta_{QED}(a_{s},a),
\]
where the beta functions can be expanded to the appropriate order, as follows
\begin{equation}
\beta_{QCD}(a_{s},a)=-\beta_{0}^{(\alpha_s)}a_{s}^{2}-\beta_{1}^{(\alpha_s \alpha)}a_{s}^{2} a-\beta_{1}^{(\alpha_s^2)}a_{s}^{3}+\mathcal{O}(a_{s}^{4}),\label{eq:36}
\end{equation}
\begin{equation}
\beta_{QED}(a_{s},a)=-\beta_{0}^{(\alpha)}a^{2}-\beta_{1}^{(\alpha \alpha_s)}a^{2} a_s-\beta_{1}^{(\alpha^2)}a^{3}+\mathcal{O}(a^{4}),\label{eq:37}
\end{equation}
where
\begin{align*}
\beta_{1}^{(\alpha_{s}\alpha)} & =-2\stackrel[i=1]{n_{f}}{\sum}e_{i}^{2}\\
\beta_{1}^{(\alpha\alpha_{s})} & =-\frac{16}{3}N_{c}\stackrel[i=1]{n_{f}}{\sum}e_{i}^{2}\\
\beta_{1}^{(\alpha^{2})} & =-4(n_{l}+N_{c}\stackrel[i=1]{n_{f}}{\sum}e_{i}^{2})
\end{align*}
Now, the splitting function matrix that determines the evolution of the singlet PDFs in NLO QCD and NLO QED approximations are as follows
\begin{equation}
P(N,Q^{2})=a_{s}P_{0}^{(1,0)}(N)+a_{s}^{2}P^{(2,0)}(N)+a P_{0}^{(0,1)}(N)+a_{s} a P^{(1,1)}(N)+a^2 P^{(0,2)}(N)\label{eq:38}
\end{equation}
where the pure QCD term is given by
\begin{equation}
\tilde{P}(N,Q^{2})=a_{s}P_{0}^{(1,0)}(N)+a_{s}^{2}P^{(2,0)}(N)+\mathcal{O}(a_{s}^{3})\label{eq:39}
\end{equation}
and the terms including the QED coupling are given by
\begin{equation}
\bar{P}(N,Q^{2})=a P_{0}^{(0,1)}(N)+a_{s} a P^{(1,1)}(N)+a^2 P^{(0,2)}(N)+\mathcal{O}(a^{3})\label{eq:40}
\end{equation}
The evolution kernels $P^{(1,1)}$ and $P^{(0,2)}$ are terms of order $\mathcal{O}(a_{s}a)$ and $\mathcal{O}(a^{2})$, respectively, that add to the DGLAP evolution equations. These evolution matrices are given by
\begin{equation}
P^{(1,1)}(N)=
\left(
\begin{array}{cccc}
\eta^+ p^{+(1,1)} & \eta^- p^{+(1,1)}& 2 \delta_e^2 p_{qg}^{(1,1)} & 2 \delta_e^2 p_{q\gamma}^{(1,1)} \\
\eta^- p^{+(1,1)}  & \eta^+ p^{+(1,1)}  & 2 \Sigma_e^2 p_{qg}^{(1,1)} & 2 \Sigma_e^2 p_{q\gamma}^{(1,1)} \\
 \eta^- p_{gq}^{(1,1)}  & \eta^+ p_{gq}^{(1,1)} & e_\Sigma^2 p_{gg}^{(1,1)} &
  e_\Sigma^2 p_{g\gamma}^{(1,1)} \\
 \eta^- p_{\gamma q}^{(1,1)}  & \eta^+ p_{\gamma q}^{(1,1)}  & e_\Sigma^2 p_{\gamma g}^{(1,1)} & e_\Sigma^2 p_{\gamma \gamma}^{(1,1)} \\
\end{array}
\right)
\label{eq:41}
\end{equation}
\begin{equation}
P^{(0,2)}(N)=\frac{1}{2} \left(
\begin{array}{cccc}
 e_u^4 p_{uu}^{+(0,2)}+ e_d^4 p_{dd}^{+(0,2)}+2 \eta^{+} \delta_e^2 p_{qq}^{S(0,2)} & e_u^4 p_{uu}^{+(0,2)}- e_d^4 p_{dd}^{+(0,2)}+2 \eta^{-} \delta_e^2 p_{qq}^{S(0,2)} & 0 & 4 \delta_e^4 p_{q \gamma}^{(0,2)}\\
 e_u^4 p_{uu}^{+(0,2)}- e_d^4 p_{dd}^{+(0,2)}+2 \eta^{-} \Sigma_e^2 p_{qq}^{S(0,2)} &  e_u^4 p_{uu}^{+{0,2}}+ e_d^4 p_{dd}^{+{(0,2)}}+2 \eta^{+} \Sigma_e^2 p_{qq}^{S(0,2)} & 0 & 4 \Sigma_e^4 p_{q \gamma}^{(0,2)}\\
 0 & 0 & 0 & 0 \\ e_u^4 p_{\gamma u}^{(0,2)}- e_d^4 p{\gamma d}^{(0,2)} & e_u^4 p_{\gamma u}^{(0,2)}+ e_d^4 p{\gamma d}^{(0,2)} & 0 & 2 e_\Sigma^4 p_{\gamma \gamma}^{(0,2)} \\
\end{array}
\right)
\label{eq:42}
\end{equation}
where
\begin{align*}
e_{\varSigma}^{4} & \equiv N_{c}(n_{u}e_{u}^{4}+n_{d}^{4}e_{d}^{4})\\
\delta_{e}^{4} & \equiv N_{c}(n_{u}e_{u}^{4}-n_{d}^{4}e_{d}^{4})
\end{align*}
The splitting functions corresponding to these evolution matrices can be found in Refs.~\cite{deFlorian:2015ujt,deFlorian:2016gvk}.
\\
In the following, for calculating the NLO QCD evolution matrix, $E_{QCD}^{N}(a_{s},a_{s0})$, we used the method represented in subsection (2.1).
\\
Now, we consider the series expansion for splitting functions Eq.~\eqref{eq:38} and for the beta function Eq.~\eqref{eq:36} and sort everything in a power series in $a_{s}$ and $a$, yields
\begin{equation}
\frac{d}{dt}E_{QCD}(t)=\left[P_{0}^{(1,0)}(N)+a_{s}(P^{(2,0)}(N)-\frac{\beta_{1}^{(\alpha_s^2)}}{\beta_0^{(\alpha_s)}}P_{0}^{(1,0)}(N))-a\frac{\beta_{1}^{(\alpha_s \alpha)}}{\beta_0^{(\alpha_s)}}P_{0}^{(1,0)}(N)\right]E_{QCD}(t)\label{eq:43}
\end{equation} 
Notice that in order to solve the above equation, we expand $a$ in terms of $a_s$. Then we can choose $a=A+B a_s+\mathcal{O}(a_s^2)$.
\\
Therefore, we can rewrite Eq.~\eqref{eq:43} as follows
\begin{equation}
\frac{d}{dt}E_{QCD}(t)=\left[(1- A\frac{\beta_{1}^{(\alpha_s \alpha)}}{\beta_0^{(\alpha_s)}} )P_{0}^{(1,0)}(N)+a_{s}(P^{(2,0)}(N)-(\frac{\beta_{1}^{(\alpha_s^2)}}{\beta_0^{(\alpha_s)}}+B\frac{\beta_{1}^{(\alpha_s \alpha)}}{\beta_0^{(\alpha_s)}})P_{0}^{(1,0)}(N))\right]E_{QCD}(t)\label{eq:44}
\end{equation} 
Introducing futher the abbreviations
\begin{equation}
R_0=(1- A\frac{\beta_{1}^{(\alpha_s \alpha)}}{\beta_0^{(\alpha_s)}} )P_{0}^{(1,0)}(N), R_1=P^{(2,0)}(N)-(\frac{\beta_{1}^{(\alpha_s^2)}}{\beta_0^{(\alpha_s)}}+B\frac{\beta_{1}^{(\alpha_s \alpha)}}{\beta_0^{(\alpha_s)}})P_{0}^{(1,0)}(N)
\end{equation} 
Finally, the solution of equation ~\eqref{eq:44} at NLO QCD and QED approximations with the method represent in subsection(2.1) is given by
\begin{eqnarray}
E_{QCD}^{N}(a_{s},a_{s0})=\{(\frac{a_{s}}{a_{s0}})^{R_0\frac{\lambda_{1}}{ \beta_{0}}}(e_{1}-\frac{(\text{\ensuremath{a_{s0}-a_{s})}}}{\text{\ensuremath{\beta_{0}}}}e_{1}R_1e_{1}-\frac{1}{\beta_{0}+\lambda_{2}-\lambda_{1}}(a_{s}(\frac{a_{s}}{a_{s0}})^{(\lambda_{2}-\lambda_{1})/\beta_{0}}-a_{s0})e_{1}R_1e_{2}\cr
-\frac{1}{\beta_{0}+\lambda_{3}-\lambda_{1}}(a_{s}(\frac{a_{s}}{a_{s0}})^{(\lambda_{3}-\lambda_{1})/\beta_{0}}-a_{s0})e_{1}R_1e_{3}-\frac{1}{\beta_{0}+\lambda_{4}-\lambda_{1}}(a_{s}(\frac{a_{s}}{a_{s0}})^{(\lambda_{4}-\lambda_{1})/\beta_{0}}-a_{s0})e_{1}R_1e_{4})\cr
+(\frac{a_{s}}{a_{s0}})^{R_0\frac{\lambda_{2}}{\beta_{0}}}(e_{2}-\frac{(\text{\ensuremath{a_{s0}-a_{s})}}}{\text{\ensuremath{\beta_{0}}}}e_{2}R_1e_{2}-\frac{1}{\beta_{0}+\lambda_{1}-\lambda_{2}}(a_{s}(\frac{a_{s}}{a_{s0}})^{(\lambda_{1}-\lambda_{2})/\beta_{0}}-a_{s0})e_{2}R_1e_{1}\cr
-\frac{1}{\beta_{0}+\lambda_{3}-\lambda_{2}}(a_{s}(\frac{a_{s}}{a_{s0}})^{(\lambda_{3}-\lambda_{2})/\beta_{0}}-a_{s0})e_{2}R_1e_{3}-\frac{1}{\beta_{0}+\lambda_{4}-\lambda_{2}}(a_{s}(\frac{a_{s}}{a_{s0}})^{(\lambda_{4}-\lambda_{2})/\beta_{0}}-a_{s0})e_{2}R_1e_{4})\cr
+(\frac{a_{s}}{a_{s0}})^{R_0\frac{\lambda_{3}}{\beta_{0}}}(e_{3}-\frac{(\text{\ensuremath{a_{s0}-a_{s})}}}{\text{\ensuremath{\beta_{0}}}}e_{3}R_1e_{3}-\frac{1}{\beta_{0}+\lambda_{1}-\lambda_{3}}(a_{s}(\frac{a_{s}}{a_{s0}})^{(\lambda_{1}-\lambda_{3})/\beta_{0}}-a_{s0})e_{3}R_1e_{1}\cr
-\frac{1}{\beta_{0}+\lambda_{2}-\lambda_{3}}(a_{s}(\frac{a_{s}}{a_{s0}})^{(\lambda_{2}-\lambda_{3})/\beta_{0}}-a_{s0})e_{3}R_1e_{2}-\frac{1}{\beta_{0}+\lambda_{4}-\lambda_{3}}(a_{s}(\frac{a_{s}}{a_{s0}})^{(\lambda_{4}-\lambda_{3})/\beta_{0}}-a_{s0})e_{3}R_1e_{4})\cr
+(\frac{a_{s}}{a_{s0}})^{R_0\frac{\lambda_{4}}{\beta_{0}}}(e_{4}-\frac{(\text{\ensuremath{a_{s0}-a_{s})}}}{\text{\ensuremath{\beta_{0}}}}e_{4}R_1e_{4}-\frac{1}{\beta_{0}+\lambda_{1}-\lambda_{4}}(a_{s}(\frac{a_{s}}{a_{s0}})^{(\lambda_{1}-\lambda_{4})/\beta_{0}}-a_{s0})e_{4}R_1e_{1}\cr
-\frac{1}{\beta_{0}+\lambda_{2}-\lambda_{4}}(a_{s}(\frac{a_{s}}{a_{s0}})^{(\lambda_{2}-\lambda_{4})/\beta_{0}}-a_{s0})e_{4}R_1e_{2}-\frac{1}{\beta_{0}+\lambda_{3}-\lambda_{4}}(a_{s}(\frac{a_{s}}{a_{s0}})^{(\lambda_{3}-\lambda_{4})/\beta_{0}}-a_{0s})e_{4}R_1e_{3})\}\label{eq:45}
\end{eqnarray} 
The QED part of evolution matrix satisfy in the following equation,
\begin{equation}
\frac{d}{d\acute{t}}E_{QED}(\acute{t})=\left[P_{0}^{(0,1)}(N)+a(P^{(1,1)}(N)-\frac{\beta_{1}^{(\alpha \alpha_s)}}{\beta_0^{(\alpha)}}P_{0}^{(0,1)}(N))+a_{s}(P^{(0,2)}(N)-\frac{\beta_{1}^{(\alpha^2)}}{\beta_0^{(\alpha)}}P_{0}^{(0,1)}(N))\right]E_{QED}(\acute{t})\label{eq:46}
\end{equation} 
where, the $\acute{t}$ parameter is $\frac{1}{\beta_{0}^{(\alpha)}} log\frac{a(Q_{0}^{2})}{a(Q^{2})}$.
In this case, we expand $a_s$ in terms of $a$. Therefore we choose $a_s=Ca+\mathcal{O}(a^2)$. We ignore the last terms in this expansion, because we work in NLO QED approximation.
We use this expansion and substitute it in equation ~\eqref{eq:46}, then we have
\begin{equation}
\frac{d}{d\acute{t}}E_{QED}(\acute{t})=\left[P_{0}^{(0,1)}(N)+a(CP^{(1,1)}(N)-C\frac{\beta_{1}^{(\alpha \alpha_s)}}{\beta_0^{(\alpha)}}P_{0}^{(0,1)}(N)+P^{(0,2)}(N)-\frac{\beta_{1}^{(\alpha^2)}}{\beta_0^{(\alpha)}}P_{0}^{(0,1)}(N))\right]E_{QED}(\acute{t})\label{eq:47}
\end{equation} 
Now, in the same method with QCD part, we define the new following parameter 
\begin{equation}
\tilde{R}_1=CP^{(1,1)}(N)-C\frac{\beta_{1}^{(\alpha \alpha_s)}}{\beta_0^{(\alpha)}}P_{0}^{(0,1)}(N)+P^{(0,2)}(N)-\frac{\beta_{1}^{(\alpha^2)}}{\beta_0^{(\alpha)}}P_{0}^{(0,1)}(N)\label{eq:48}
\end{equation} 
Therefore, the QED evolution matrix obtain in the following form,
\begin{eqnarray}
E_{QED}^{N}(a,a_{0})=\{(\frac{a}{a_{0}})^{\frac{\tilde{\lambda}_{1}}{ \beta^{(\alpha)}_{0}}}(\tilde{e}_{1}-\frac{(\text{\ensuremath{a_{0}-a)}}}{\text{\ensuremath{\beta^{(\alpha)}_{0}}}}\tilde{e}_{1}\tilde{R}_1\tilde{e}_{1}-\frac{1}{\beta^{(\alpha)}_{0}+\tilde{\lambda}_{2}-\tilde{\lambda}_{1}}(a(\frac{a}{a_{0}})^{(\tilde{\lambda}_{2}-\tilde{\lambda}_{1})/\beta^{(\alpha)}_{0}}-a_{0})\tilde{e}_{1}\tilde{R}_1\tilde{e}_{2}\cr
-\frac{1}{\beta^{(\alpha)}_{0}+\tilde{\lambda}_{3}-\tilde{\lambda}_{1}}(a(\frac{a}{a_{0}})^{(\tilde{\lambda}_{3}-\tilde{\lambda}_{1})/\beta^{(\alpha)}_{0}}-a_{0})\tilde{e}_{1}\tilde{R}_1\tilde{e}_{3}-\frac{1}{\beta^{(\alpha)}_{0}+\tilde{\lambda}_{4}-\tilde{\lambda}_{1}}(a(\frac{a}{a_{0}})^{(\tilde{\lambda}_{4}-\tilde{\lambda}_{1})/\beta^{(\alpha)}_{0}}-a_{0})\tilde{e}_{1}\tilde{R}_1\tilde{e}_{4})\cr
+(\frac{a}{a_{0}})^{\frac{\tilde{\lambda}_{2}}{\beta^{(\alpha)}_{0}}}(\tilde{e}_{2}-\frac{(\text{\ensuremath{a_{0}-a)}}}{\text{\ensuremath{\beta^{(\alpha)}_{0}}}}\tilde{e}_{2}\tilde{R}_1\tilde{e}_{2}-\frac{1}{\beta^{(\alpha)}_{0}+\tilde{\lambda}_{1}-\tilde{\lambda}_{2}}(a(\frac{a}{a_{0}})^{(\tilde{\lambda}_{1}-\tilde{\lambda}_{2})/\beta^{(\alpha)}_{0}}-a_{0})\tilde{e}_{2}\tilde{R}_1\tilde{e}_{1}\cr
-\frac{1}{\beta^{(\alpha)}_{0}+\tilde{\lambda}_{3}-\tilde{\lambda}_{2}}(a(\frac{a}{a_{0}})^{(\tilde{\lambda}_{3}-\tilde{\lambda}_{2})/\beta^{(\alpha)}_{0}}-a_{0})\tilde{e}_{2}\tilde{R}_1\tilde{e}_{3}-\frac{1}{\beta^{(\alpha)}_{0}+\tilde{\lambda}_{4}-\tilde{\lambda}_{2}}(a(\frac{a}{a_{0}})^{(\tilde{\lambda}_{4}-\tilde{\lambda}_{2})/\beta^{(\alpha)}_{0}}-a_{0})\tilde{e}_{2}\tilde{R}_1\tilde{e}_{4})\cr
+(\frac{a}{a_{0}})^{\frac{\tilde{\lambda}_{3}}{\beta^{(\alpha)}_{0}}}(\tilde{e}_{3}-\frac{(\text{\ensuremath{a_{0}-a)}}}{\text{\ensuremath{\beta^{(\alpha)}_{0}}}}\tilde{e}_{3}\tilde{R}_1\tilde{e}_{3}-\frac{1}{\beta^{(\alpha)}_{0}+\tilde{\lambda}_{1}-\tilde{\lambda}_{3}}(a(\frac{a}{a_{0}})^{(\tilde{\lambda}_{1}-\tilde{\lambda}_{3})/\beta^{(\alpha)}_{0}}-a_{0})\tilde{e}_{3}\tilde{R}_1\tilde{e}_{1}\cr
-\frac{1}{\beta^{(\alpha)}_{0}+\tilde{\lambda}_{2}-\tilde{\lambda}_{3}}(a(\frac{a}{a_{0}})^{(\tilde{\lambda}_{2}-\tilde{\lambda}_{3})/\beta^{(\alpha)}_{0}}-a_{0})\tilde{e}_{3}\tilde{R}_1\tilde{e}_{2}-\frac{1}{\beta^{(\alpha)}_{0}+\tilde{\lambda}_{4}-\tilde{\lambda}_{3}}(a(\frac{a}{a_{0}})^{(\tilde{\lambda}_{4}-\tilde{\lambda}_{3})/\beta^{(\alpha)}_{0}}-a_{0})\tilde{e}_{3}\tilde{R}_1\tilde{e}_{4})\cr
+(\frac{a}{a_{0}})^{\frac{\tilde{\lambda}_{4}}{\beta^{(\alpha)}_{0}}}(\tilde{e}_{4}-\frac{(\text{\ensuremath{a_{0}-a)}}}{\text{\ensuremath{\beta^{(\alpha)}_{0}}}}\tilde{e}_{4}\tilde{R}_1\tilde{e}_{4}-\frac{1}{\beta^{(\alpha)}_{0}+\tilde{\lambda}_{1}-\tilde{\lambda}_{4}}(a(\frac{a}{a_{0}})^{(\tilde{\lambda}_{1}-\tilde{\lambda}_{4})/\beta^{(\alpha)}_{0}}-a_{0})\tilde{e}_{4}\tilde{R}_1\tilde{e}_{1}\cr
-\frac{1}{\beta^{(\alpha)}_{0}+\tilde{\lambda}_{2}-\tilde{\lambda}_{4}}(a(\frac{a}{a_{0}})^{(\tilde{\lambda}_{2}-\tilde{\lambda}_{4})/\beta^{(\alpha)}_{0}}-a_{0})\tilde{e}_{4}\tilde{R}_1\tilde{e}_{2}-\frac{1}{\beta^{(\alpha)}_{0}+\tilde{\lambda}_{3}-\tilde{\lambda}_{4}}(a(\frac{a}{a_{0}})^{(\tilde{\lambda}_{3}-\tilde{\lambda}_{4})/\beta^{(\alpha)}_{0}}-a_{0s})\tilde{e}_{4}\tilde{R}_1\tilde{e}_{3})\}
\label{eq:49}
\end{eqnarray} 
We calculated the values of fixed parameters A, B and C for the different active flavour numbers. We listed them in Table ~\eqref{table2}.
\begin{table}[H]
\centering{}\caption{The parameters A, B and C for the different values of $n_{f}$}
\begin{tabular}{cccc}
\hline 
$n_{f}$ & A & B & C\tabularnewline
\hline 
\hline 
3 & 0.000605 & -0.000199 & -0.740\tabularnewline
\hline 
4 & 0.000613 & -0.000443 & -1.013\tabularnewline
\hline 
5 & 0.000617 & -0.000548 & -1.159\tabularnewline
\hline 
\end{tabular}\label{table2}
\end{table}

\subsection{The non-singlet PDFs with NLO QED corrections}
In the non-singlet case, the evolution equations are scalar, so that the
commutation relations in Eq. ~\eqref{eq:20} are vanished.  In the same procedure established in subsection (2.2), we calculate the evolution operators for QCD part in the next to leading order QCD and QED approximations for the non-singlet case. These QCD evolution operators for the non-singlet PDFs introduced in Eq. ~\eqref{eq:5} are listed in Table ~\eqref{table3}.
\\
\begin{table}[H]
\centering{}\caption{The QCD evolution operators, $E_{QCD}^{N}(Q^{2},Q_{0}^{2})$, for the non-singlet
PDFs with NLO QED corrections}
\begin{tabular}{c>{\centering}m{12cm}}
\hline
\textbf{$\mathbf{\mathit{f}_{i}}$} & $E_{QCD}^{N}(Q^{2},Q_{0}^{2})$ \tabularnewline
\hline
$f_{5}$ & \begin{singlespace}
\centering{}$(1+\frac{a_{s}-a_{s0}}{\beta_{0}^{(\alpha_s)}}(P^{-(2,0)}_{qq}(N)-(\frac{\beta_{1}^{(\alpha_s^2)}}{\beta_0^{(\alpha_s)}}+B\frac{\beta_{1}^{(\alpha_s \alpha)}}{\beta_0^{(\alpha_s)}})P^{(0)}_{qq}(N)))(\frac{a_{s}}{a_{s0}})^{\frac{(1- A\frac{\beta_{1}^{(\alpha_s \alpha)}}{\beta_0^{(\alpha_s)}} )P^{(0)}_{qq}(N)}{\beta_{0}^{(\alpha_s)}}}$
\end{singlespace}
\tabularnewline
\hline
$f_{6}$ & \begin{singlespace}
\centering{}$(1+\frac{a_{s}-a_{s0}}{\beta_{0}^{(\alpha_s)}}(P^{-(2,0)}_{qq}(N)-(\frac{\beta_{1}^{(\alpha_s^2)}}{\beta_0^{(\alpha_s)}}+B\frac{\beta_{1}^{(\alpha_s \alpha)}}{\beta_0^{(\alpha_s)}})P^{(0)}_{qq}(N)))(\frac{a_{s}}{a_{s0}})^{\frac{(1- A\frac{\beta_{1}^{(\alpha_s \alpha)}}{\beta_0^{(\alpha_s)}} )P^{(0)}_{qq}(N)}{\beta_{0}^{(\alpha_s)}}}$
\end{singlespace}
\tabularnewline
\hline
$f_{7}$ & \begin{singlespace}
\centering{}$(1+\frac{a_{s}-a_{s0}}{\beta_{0}^{(\alpha_s)}}(P^{+(2,0)}_{qq}(N)-(\frac{\beta_{1}^{(\alpha_s^2)}}{\beta_0^{(\alpha_s)}}+B\frac{\beta_{1}^{(\alpha_s \alpha)}}{\beta_0^{(\alpha_s)}})P^{(0)}_{qq}(N)))(\frac{a_{s}}{a_{s0}})^{\frac{(1- A\frac{\beta_{1}^{(\alpha_s \alpha)}}{\beta_0^{(\alpha_s)}} )P^{(0)}_{qq}(N)}{\beta_{0}^{(\alpha_s)}}}$
\end{singlespace}
\tabularnewline
\hline
$f_{8}$ & \begin{singlespace}
\centering{}$(1+\frac{a_{s}-a_{s0}}{\beta_{0}^{(\alpha_s)}}(P^{+(2,0)}_{qq}(N)-(\frac{\beta_{1}^{(\alpha_s^2)}}{\beta_0^{(\alpha_s)}}+B\frac{\beta_{1}^{(\alpha_s \alpha)}}{\beta_0^{(\alpha_s)}})P^{(0)}_{qq}(N)))(\frac{a_{s}}{a_{s0}})^{\frac{(1- A\frac{\beta_{1}^{(\alpha_s \alpha)}}{\beta_0^{(\alpha_s)}} )P^{(0)}_{qq}(N)}{\beta_{0}^{(\alpha_s)}}}$
\end{singlespace}
\tabularnewline
\hline
$f_{9}$ & \begin{singlespace}
\centering{}$(1+\frac{a_{s}-a_{s0}}{\beta_{0}^{(\alpha_s)}}(P^{+(2,0)}_{qq}(N)-(\frac{\beta_{1}^{(\alpha_s^2)}}{\beta_0^{(\alpha_s)}}+B\frac{\beta_{1}^{(\alpha_s \alpha)}}{\beta_0^{(\alpha_s)}})P^{(0)}_{qq}(N)))(\frac{a_{s}}{a_{s0}})^{\frac{(1- A\frac{\beta_{1}^{(\alpha_s \alpha)}}{\beta_0^{(\alpha_s)}} )P^{(0)}_{qq}(N)}{\beta_{0}^{(\alpha_s)}}}$
\end{singlespace}
\tabularnewline
\hline
\end{tabular}\label{table3}
\end{table}
With the same procedure, we can obtain the QED evolution operators for the non-singlet PDFs in Eq. ~\eqref{eq:5} with NLO QED corrections. We give these operators in table ~\eqref{table4}. 
\begin{table}[H]
\centering{}\caption{The QED evolution operators, $E_{QED}^{N}(Q^{2},Q_{0}^{2})$, for the non-singlet
PDFs with NLO QED corrections}
\begin{tabular}{c>{\centering}m{13cm}}
\hline
\textbf{$\mathbf{\mathit{f}_{i}}$} & $E_{QED}^{N}(Q^{2},Q_{0}^{2})$ \tabularnewline
\hline
$f_{5}$ & \begin{singlespace}
\centering{}$(1+\frac{a-a_{0}}{\beta_{0}^{(\alpha)}}(C e_d^2 p^{-(1,1)}(N)-C \frac{\beta_{1}^{(\alpha \alpha_s)}}{\beta_0^{(\alpha)}}\tilde{p}^{(0)}_{qq}(N)+e_d^4 p^{-(0,2)}_{qq}(N)-\frac{\beta_{1}^{(\alpha^2)}}{\beta_0^{(\alpha)}}\tilde{p}^{(0)}_{qq}(N))(\frac{a}{a_{0}})^{e_d^2\frac{\tilde{p}^{(0)}_{qq}(N)}{\beta_{0}^{(\alpha)}}}$
\end{singlespace}
\tabularnewline
\hline
$f_{6}$ & \begin{singlespace}
\centering{}$(1+\frac{a-a_{0}}{\beta_{0}^{(\alpha)}}(C e_u^2 p^{-(1,1)}(N)-C \frac{\beta_{1}^{(\alpha \alpha_s)}}{\beta_0^{(\alpha)}}\tilde{p}^{(0)}_{qq}(N)+e_u^4 p^{-(0,2)}_{qq}(N)-\frac{\beta_{1}^{(\alpha^2)}}{\beta_0^{(\alpha)}}\tilde{p}^{(0)}_{qq}(N))(\frac{a}{a_{0}})^{e_u^2\frac{\tilde{p}^{(0)}_{qq}(N)}{\beta_{0}^{(\alpha)}}}$
\end{singlespace}
\tabularnewline
\hline
$f_{7}$ & \begin{singlespace}
\centering{}$(1+\frac{a-a_{0}}{\beta_{0}^{(\alpha)}}(C e_d^2 p^{+(1,1)}(N)-C \frac{\beta_{1}^{(\alpha \alpha_s)}}{\beta_0^{(\alpha)}}\tilde{p}^{(0)}_{qq}(N)+e_d^4 p^{+(0,2)}_{qq}(N)-\frac{\beta_{1}^{(\alpha^2)}}{\beta_0^{(\alpha)}}\tilde{p}^{(0)}_{qq}(N))(\frac{a}{a_{0}})^{e_d^2\frac{\tilde{p}^{(0)}_{qq}(N)}{\beta_{0}^{(\alpha)}}}$
\end{singlespace}
\tabularnewline
\hline
$f_{8}$ & \begin{singlespace}
\centering{}$(1+\frac{a-a_{0}}{\beta_{0}^{(\alpha)}}(C e_u^2 p^{+(1,1)}(N)-C \frac{\beta_{1}^{(\alpha \alpha_s)}}{\beta_0^{(\alpha)}}\tilde{p}^{(0)}_{qq}(N)+e_u^4 p^{+(0,2)}_{qq}(N)-\frac{\beta_{1}^{(\alpha^2)}}{\beta_0^{(\alpha)}}\tilde{p}^{(0)}_{qq}(N))(\frac{a}{a_{0}})^{e_u^2\frac{\tilde{p}^{(0)}_{qq}(N)}{\beta_{0}^{(\alpha)}}}$
\end{singlespace}
\tabularnewline
\hline
$f_{9}$ & \begin{singlespace}
\centering{}$(1+\frac{a-a_{0}}{\beta_{0}^{(\alpha)}}(C e_d^2 p^{+(1,1)}(N)-C \frac{\beta_{1}^{(\alpha \alpha_s)}}{\beta_0^{(\alpha)}}\tilde{p}^{(0)}_{qq}(N)+e_d^4 p^{+(0,2)}_{qq}(N)-\frac{\beta_{1}^{(\alpha^2)}}{\beta_0^{(\alpha)}}\tilde{p}^{(0)}_{qq}(N))(\frac{a}{a_{0}})^{e_d^2\frac{\tilde{p}^{(0)}_{qq}(N)}{\beta_{0}^{(\alpha)}}}$
\end{singlespace}
\tabularnewline
\hline
\end{tabular}\label{table4}
\end{table}

\section{Test the accuracy of the solutions}

In the section (2), we analytically solved the QCD$\otimes$QED DGLAP evolution equations in NLO QCD and LO QED approximations. We obtain the singlet and non-singlet parton distribution functions with QED corrections in x space and for $Q^{2}>Q_{0}^{2} (GeV^{2})$.

The main question here is: How we can check our analytical solutions? The best way to be sure about the correctness of these solutions is to take the initial PDFs from  a public code such as the CT14QED global analysis code \cite{Schmidt:2015zda}
at a fixed scale of $Q_{0}^2$  and then we can calculate all of the parton distribution functions inside proton at some values of $Q^{2}$.  In order to do this, we run the CT14QED program \cite{Schmidt:2015zda} with  $P_0^ {\gamma } \leq 14   \% $ for the inelastic photon PDF in initial scale of $Q_0= 1.295 GeV$.  Then, we evolved this initial PDFs with current solutions for the QCD$\otimes$QED DGLAP evolution equations up to some values of $Q^{2}$. These PDFs are determined for the central QCD coupling of $\alpha_s(M_Z) = 0.118$.

We perform the evolution equations in a fixed flavour number (FFN) scheme for all of the distributions.  Since in this article we consider only five active flavours. The lowest order QED beta function in this case takes the value
\begin{equation}
\beta_{0}^{(\alpha)}=-\frac{80}{9}
\end{equation}
where we suppose $\mu=1.777GeV$, then $a(\mu^{2})=\frac{1}{4\pi}\frac{1}{133.4}$  in equation. ~\eqref{eq:10}.

We also assumed the symmetry between quarks- anti quarks distributions. Then the corresponding valence distributions vanish and we have $s= \bar s$, $c= \bar c$ and $b= \bar b$.

\begin{figure}
\begin{centering}
\includegraphics[scale=0.45]{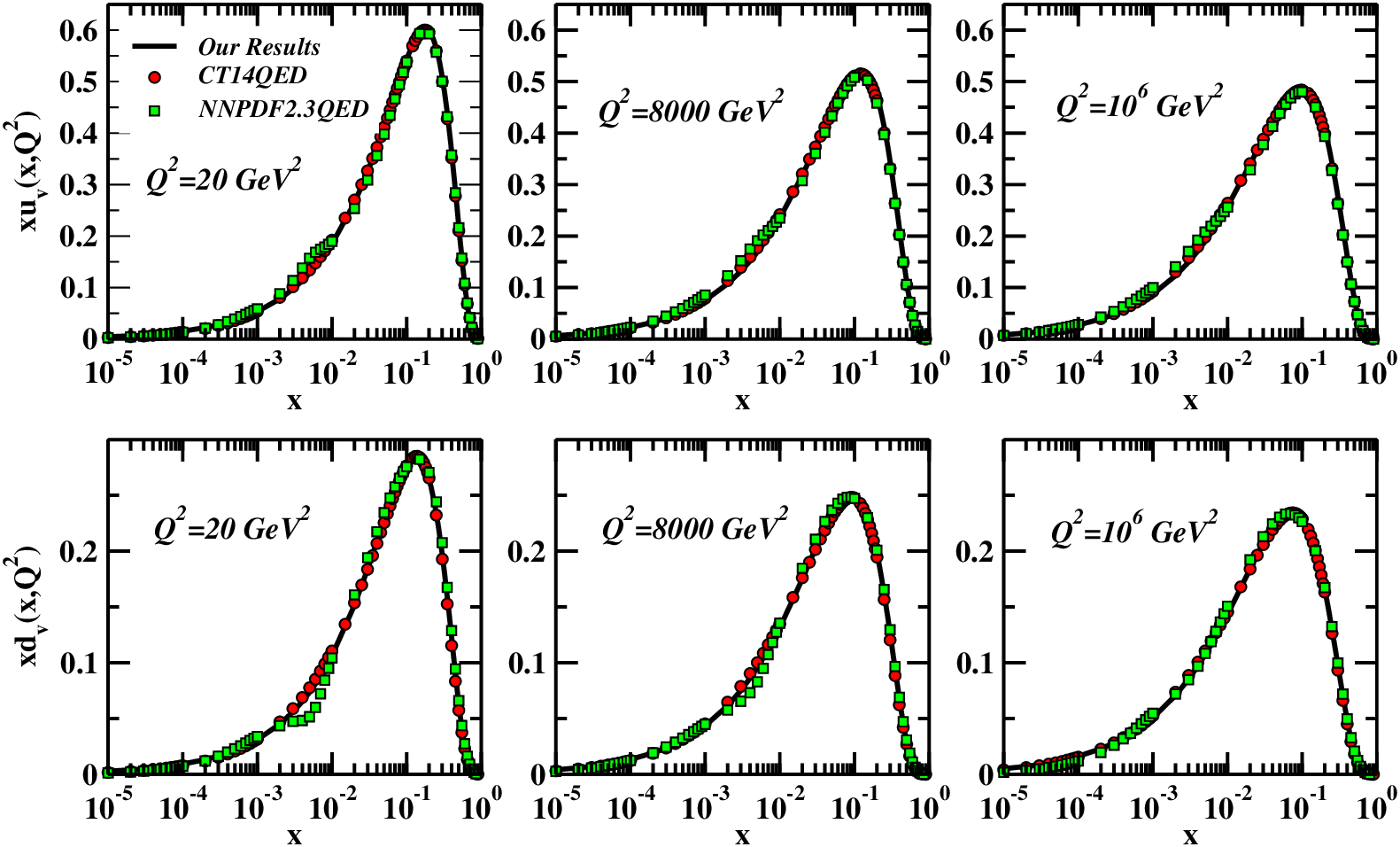}
\par\end{centering}
\caption{The parton distribution $xu_{v}(x,Q^{2})$ and $xd_{v}(x,Q^{2})$
at the different values of $Q^{2}$. The solid line is our model,
square scatter is the CT14QED global analysis code, circle scatter is the APFEL (NNPDF2.3QED) code. }
\label{fig1}
\end{figure}

\begin{figure}
\begin{centering}
\includegraphics[scale=0.42]{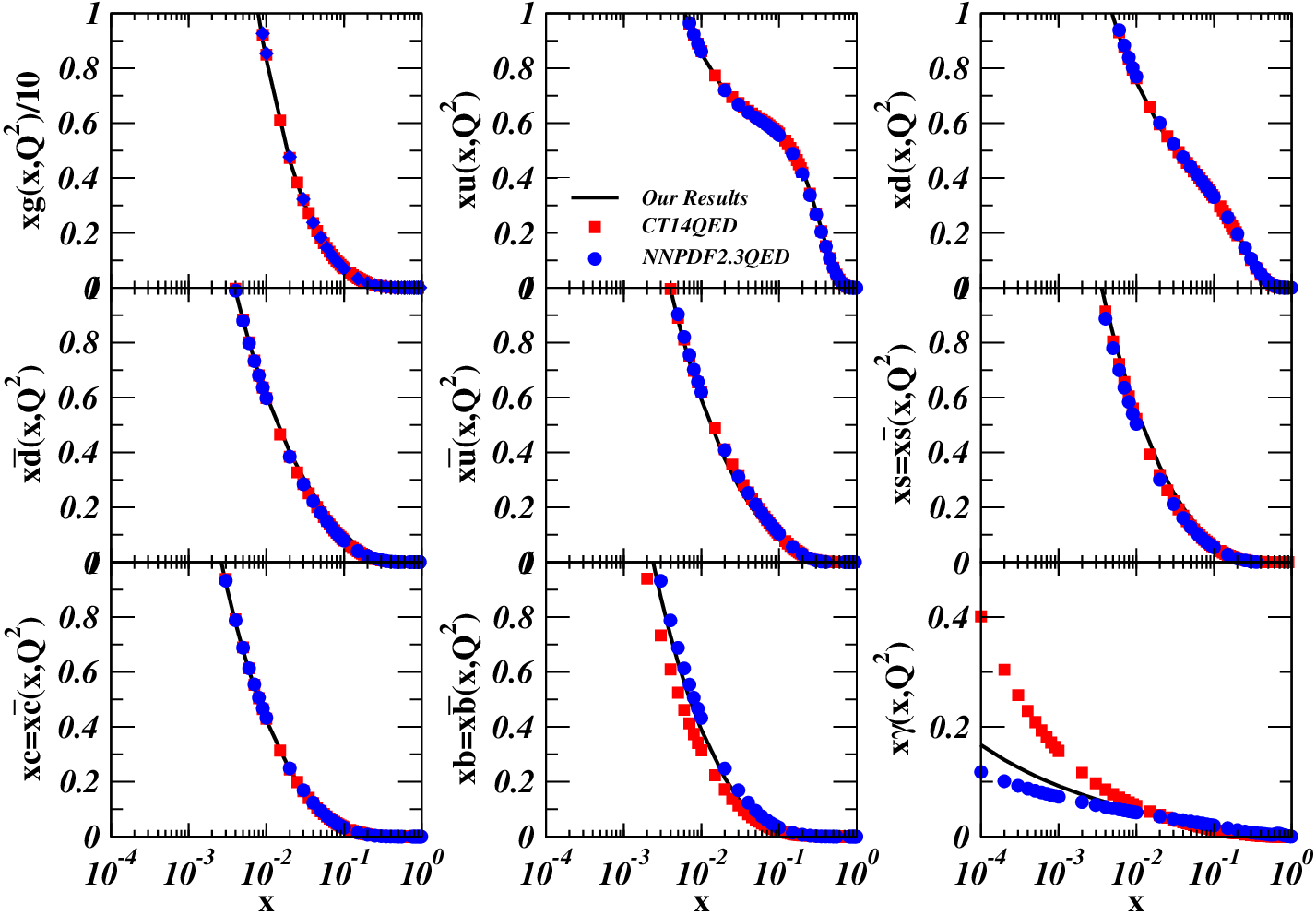}
\par\end{centering}
\caption{The parton distribution functions in comparison with the CT14QED and
APFEL (NNPDF2.3QED) global analysis codes
 at $Q^{2}=10^{6}GeV^{2}$.}

\label{fig2}
\end{figure}

\begin{figure}
\begin{centering}
\includegraphics[scale=0.4]{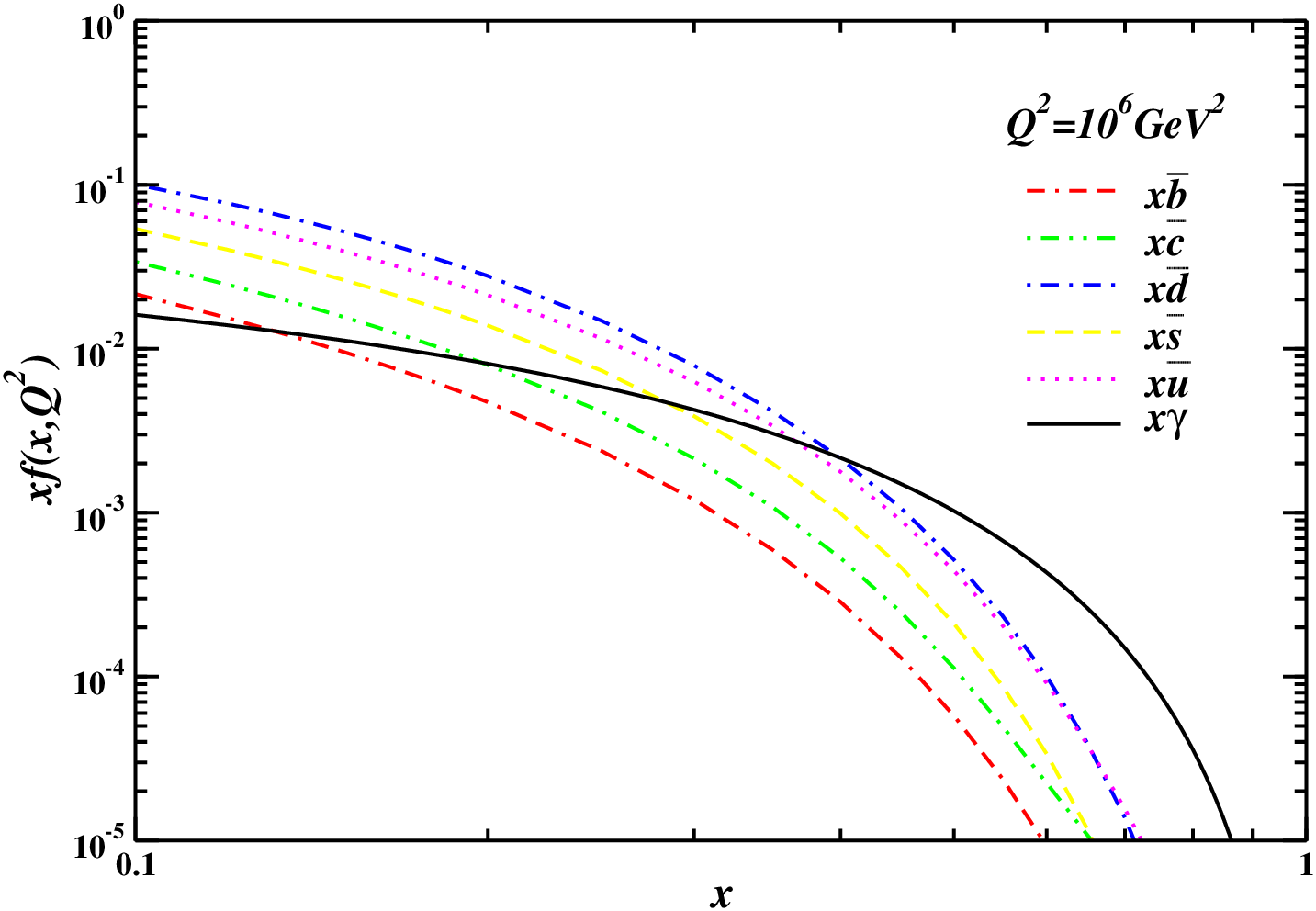}
\par\end{centering}
\caption{The Sea quarks and photon distribution functions at $Q^{2}=10^{6}GeV^{2}$
as a function of x}
\label{fig3}
\end{figure}

The results of combined QCD$\otimes$QED evolution are summarized on plots of Figs.~\eqref{fig1} to ~\eqref{fig3}, where we compare the evolution of valance quarks, sea quarks, gluon and photon PDFs with the solutions of two available codes, such as the QavD solutions implemented in the APFEL (NNPDF2.3QED) \cite{Bertone:2013vaa} program and solutions extracted from the CT14QED global analysis code.

Comparison between the APFEL (NNPDF2.3QED) predictions and the CT14QED evolution with our results for the valance quarks at different values of $Q^{2}$ is shown in Fig. ~\eqref{fig1}. An excellent agreement is found for all flavors. It is clear from Fig. ~\eqref{fig1} that  increasing  the value of $Q^{2}$ would lead to a decrease in the value of decline rate the quark valence distribution functions. This means that the presence of photon distribution function influences on the decline rate.

The sea quarks, gluon and photon distribution functions are plotted in Fig. ~\eqref{fig2}, where PDFs have been evolved from $Q_0= 1.295 GeV$ up to $Q=10^{3} GeV$ and they are compared with the CT14QED and APFEL (NNPDF2.3QED) PDF sets. It is obvious that  a good agreement is achieved.

One observation from Fig. ~\eqref{fig2} is that the sea-quark distributions , especially c and b quarks distributions,  have more impact on the photon PDF than does the initial photon distribution at high $Q^2$. These photon distribution become more significant at high $Q^2$ where more photon are produced through radiation of the quarks.

Figure ~\eqref{fig3} shows the sea quarks and photon
distribution functions in x space at scale of $Q^{2}=10^{6}GeV^{2}$.
It is worth to notice that, the photon distribution functions are
larger than the sea quarks distribution functions at high scale of
energy and for the large values of x.

In tables ~\eqref{table2} and ~\eqref{table3} we present the Patron Evolution program \cite{Roth:2004ti} results and the CT14QED code \cite{Schmidt:2015zda} results for the singlet part of the parton distribution functions for $Q=100 GeV$ and the different values of x. we show our results for the singlet part of the PDFs in the same scale energy $Q=100 GeV$ and for the several values of x, in table ~\eqref{table4}. The comparison all of the parton distribution functions in table ~\eqref{table4} with table ~\eqref{table3} show that a very good agreement is between our results and the results extracted of CT14QED code. It is clear that the QED corrections can reduce the present errors for all of the values of x. More substantial differences appear for the photon PDF, in this case the solutions differ by up to $1\%$, both at small and large values of x, however this level of agreement is still acceptable in view of the technical differences between both programs.

In conclusion, we found a good level of agreement for all of the comparison performed in this section. These guarantees that we implements correctly the QCD$\otimes$QED evolutions, therefore it can be used in any global parametrization of PDFs with  experimental data.

Our analytical solutions can be used to interpret the LHC data in a global fit parameterization. To do this, recent ATLAS measurement of high-mass Drell-yan dilepton production data combined with inclusive deep-inelastic scattering cross-section data can perform a solid determination of the proton PDFs including the photon PDF ~\cite{Giuli:2017oii}.
\begin{table*}
\caption{The singlet part of the parton distribution functions obtained from the Parton Evolution program \cite{Roth:2004ti} at $Q=100GeV$.}
\centering{}%
\begin{tabular}{cccccccc}
\hline
x & $10^{-5}$ & $10^{-4}$ & $10^{-3}$ & $10^{-2}$ & $10^{-1}$ & 0.3 & 0.7\tabularnewline
\hline
\hline
$\Delta$ & -2.017 & -1.255 & -0.74 & -0.312 & 0.125 & 0.1423 & 0.0113\tabularnewline
$\Sigma$ & 66.613 & 29.466 & 11.947 & 4.321 & 1.2078 & 0.3670 & 0.0154\tabularnewline
$ g$ & 252.004 & 100.717 & 34.863 & 9.226 & 1.0559 & 0.0963 & 0.0004\tabularnewline
$\gamma$ & 0.408 & 0.178 & 0.071 & 0.024 & 0.0042 & 0.0007 & $1\times10^{-5}$\tabularnewline
\hline
\end{tabular}
\label{table5}
\end{table*}

\begin{table*}
\caption{The CT14QED code \cite{Schmidt:2015zda} results for the singlet part of the parton distribution functions at $Q=100GeV$.}
\centering{}%
\begin{tabular}{cccccccc}
\hline
x & $10^{-5}$ & $10^{-4}$ & $10^{-3}$ & $10^{-2}$ & $10^{-1}$ & 0.3 & 0.7\tabularnewline
\hline
\hline
$\Delta$ & -12.287 & -5.228 & -1.953 & -0.578 & 0.126 & 0.1974 & 0.01226\tabularnewline
$\Sigma$ & 70.09 & 31.564 & 12.889 & 4.613 & 1.3772 & 0.501 & 0.017\tabularnewline
$ g$ & 207.8 & 84.99 & 30.08 & 8.067 & 0.882 & 0.937 & 0.000346\tabularnewline
$\gamma$ & 0.4254 & 0.2017 & 0.09187 & 0.03916 & 0.01069 & 0.002338 & $4.18\times10^{-5}$\tabularnewline
\hline
\end{tabular}
\label{table6}
\end{table*}
\begin{table*}
\caption{Our evolved singlet parton distribution functions at  $Q=100GeV$.}
\centering{}%
\begin{tabular}{cccccccc}
\hline
x & $10^{-5}$ & $10^{-4}$ & $10^{-3}$ & $10^{-2}$ & $10^{-1}$ & 0.3 & 0.7\tabularnewline
\hline
\hline
$\Delta$ & -12.756 & -5.812 & -2.403 & -0.788 & 0.098 & 0.201 &0.013\tabularnewline
$\Sigma$ &71.889 & 32.553 &13.489 & 4.871 &1.424 & 0.498 & 0.0193\tabularnewline
$ g$ & 199.708 &82.919 &29.507 &7.869&0.853 &0.0892 & 0.000334\tabularnewline
$\gamma$ &0.29765 & 0.167 & 0.09226&0.0476 & 0.01615 & 0.00425 & 0.000149 \tabularnewline
\hline
\end{tabular}
\label{table7}
\end{table*}

\section{Summary and Conclusions}
The analytical solution is performed based on the Mellin transforms
to obtain the QCD$\otimes$QED parton distribution functions. Our
calculations are done in the NLO QCD and NLO QED approximations.

To be sure about the correctness of analytical solutions, we take
the initial PDFs from the newly release CT14QED global parameterization
at a fixed scale of $Q_{0}=1.295GeV$. Therefore, here we only checked the correctness of our solutions in the NLO QCD and LO QED approximations. Then we evolved
all the parton distribution functions inside proton to some values
of $Q^{2}$. We determine the parton distribution functions  at the different values of $Q^{2}$, and compare
them with APFEL (NNPDF2.3QED) and CT14QED PDFs set. Our results
present a good agreement with them. The results show that with increasing
the value of $Q^{2}$, the contribution of valence quarks are decreased.
Also it is found that the contribution of photon distribution
function in comparison to the sea quark distribution functions is
significant, especially at the large values of $x$ and the high values of $Q^{2}$.
Briefly, the most striking points in this paper are as follows,
\begin{enumerate}
\item This method can be generalized to $N^{2}LO$ and the higher
order corrections of QCD and QED.
\item The results show that we found the exact analytical solutions for the QCD$\otimes$QED DGLAP evolution equations in
the Mellin space. Up to now less attention has been paid to such solutions
for the QCD$\otimes$QED DGLAP equations in the literature.
\item In our solutions, the QED running coupling is not constant and $Q^{2}$ dependence of this
parameter is considered.
\end{enumerate}
Last, but not least, these analytical solutions can be used to
determine the QCD$\otimes$QED parton distribution functions via
a global parameterization to the experimental data in the future.

\section*{Acknowledgments}
The authors would like to thank F. Arash for carefully reading the manuscript, fruitful discussion and critical remarks.

\end{document}